\documentclass[]{raa}
\usepackage{graphicx,times}
\usepackage{natbib}
\usepackage{amssymb,amsmath}
\usepackage{rotating}
\usepackage{longtable}
\usepackage{supertabular}
\usepackage{longtable,lscape}
\bibpunct{(}{)}{;}{a}{}{,}

\def\apj{ApJ}
\def\aap{A\&A}
\def\mnras{MNRAS}
\def\apjs{ApJS}
\def\apjl{ApJL}

\usepackage[a4paper=true,dvipdfm=true,pagebackref=true]{hyperref}
\hypersetup{pdftitle = The title of my PDF, pdfauthor = My name, pdfsubject= The subject, pdfkeywords = keyword1 keyword2 keyword3}
\hypersetup{colorlinks = true, linkcolor = green, anchorcolor = red, citecolor = blue, filecolor = red, pagecolor = red, urlcolor = red}
\UseRawInputEncoding

\begin{document}

   \title{The intra-day Optical Monitoring of BL Lacerate Object 1ES\,1218+304 at Its Highest X-ray Flux Level
$^*$
\footnotetext{\small $*$ Supported by the National Natural Science Foundation of China.}
}

 \volnopage{ {\bf 2012} Vol.\ {\bf X} No. {\bf XX}, 000--000}
   \setcounter{page}{1}

   \author{Si-Si Sun\inst{1,2}, Hua-Li Li\inst{2,3}, Xing Yang\inst{1}, Jing L\"{u}
           \inst{1}, Da-Wei Xu\inst{2,3}, Jing Wang\inst{1,2}
   }

   \institute{ Guangxi Key Laboratory for Relativistic Astrophysics, School of Physical Science and Technology, Guangxi University, Nanning
              530004, Peoples Pepublic of China; {\it sssun@bao.ac.cn}\\
        \and
              Key Laboratory of Space Astronomy and Technology, National Astronomical Observatories, Chinese Academy of Science,Beijing 100101,
              China; {\it lhl@nao.cas.cn}\\
	\and
              School of Astronomy and Space Science,University of Chinese Academy of Sciences, Beijing,China
\vs \no
   {\small Received ; accepted }
}

\abstract{
We here report a monitor  of the BL Lac object 1ES\,1218+304 in both $B$- and $R$-bands by the GWAC-F60A telescope in eight nights, when it was
triggerd to be at its highest X-ray flux in history by the VERITAS Observatory and \it Swift \rm follow-ups.
Both ANOVA and $\chi^2$-test enable us to clearly reveal an intra-day variability in optical wavelengths in seven out of the eight nights.
A bluer-when-brighter chromatic relationship has been clearly identified in five out of the eight nights, which can be well explained by the
shock-in-jet model. In addtion, a quasi-periodic oscilation phenomenon in both bands could be tentatively identified in the first night.
A positive delay between the two bands has been revealed in three out of the eight nights, and a negative one in the other nights.
The identfied minimum time delay enables us to estimate the $M_{\mathrm{BH}}=2.8\times10^7  \rm M_{\odot}$ that is invalid.
}

   \authorrunning{S.-S. Sun et al. }            
   \titlerunning{The intra-day optical variation of 1ES\,1218+304}  
   \maketitle

%
\section{Introduction}           
\label{sect:intro}


Blazars are the most extreme subclass of active galactic nuclei (AGNs) with a relativistic jet oriented at a small single towards an observer.
The jet is believed to be generated by extracting the rotation energy of the central supermassive black hole (SMBH) through either Blandford-Znajek (BZ)
mechanism \citep{blandford77} or Blandford-Payne (BP) mechanism \citep{blandford82}.
Because of the jet beaming effect,
blazars are characterized by large amplitude and rapid variability at all wavelengths, high and variable polarization, superluminal jet speeds, and
compact radio emission \citep{angel80,urry95}.
There are two types of blazars: one is flat-spectrum radio quasar (FSRQ), and another is BL Lacerate (BL Lac) object.
BL Lac object is lack of strong emission lines in its spectrum and has strong variability on different timescales, from years down to minutes \citep{poon09}.

There are two bumps in the spectral energy distributions (SED) of a blazar. The low-frequency bump, which peaks at from radio to UV/X-ray,
is produced by the synchrotron emission from the relativistic electrons in the magnetic field.
The high-frequency one from X-ray to $\gamma$-ray
is believed to be contributed from the emission of inverse Compton scattering of the low-frequency photons (e.g., \citep{ulrich97, Bottcher07, Dermer09, Zhang10}).
According to the peak frequency of the low-frequency bump,
BL Lac can be classified into three groups, namely low-, intermediate-, high-frequency peak sources (LBL, IBL and HBL).

Multi-wavelengths monitor is a powerful tool for investigating the properties of the spatially unresovled jet observed in blazars (e.g., \citep{Sambruna07, Marscher08, singh15,liao14}).
For instance, the shortest variability timescale is useful for constraining the size and location of the emitting region.
The long-term correlated variability between low and high energy emissions observed from blazars generally supports the leptonic model for the
jet (e.g., \citep{Zhang10,Zhang18}), althogh this conclusion is argued against by the detection of the TeV neutrino events (e.g., \citep{Rodrigues18, singh20}).
Previous studies indicate sophisticated spectral variability behavior of blazars, which depends on variation modes and time scales (e.g., \citep{ulrich97,
Bottcher07,Meng18}).
Both bluer-when-brighter (BWB) and redder-when-brighter (RWB) chromatisms are, in fact, detected in blazars (e.g., \citep{Vagnetti03,Bonning12,Bottcher09,wu12}).

We here report an optical monitor for BL Lac object 1ES\,1218+304 (R.A.=12$^h$21$^m$21.941$^s$, DEC=+$30^{\circ}10^{'}37.1{''}$, J2000.0) in multi-bands
in 2019 January, when the object was spotted in a flare in very-high-energy (VHE, $>100$ GeV) $\gamma$-ray by the VERITAS Observatory (e.g.,\citep{Mirzoyan19}).
The analysis of the \it Swfit \rm reveals that the X-ray Flux of the source is also in a historical flaring state with the highest flux in the range of 2-10 keV
reaching to $(1.83\pm0.14)\times10^{-10}\ \mathrm{erg\ cm^{-2}\ s^{-1}}$ on the 2019-01-06T01:31:32 which is a factor of ~5 times higher than the average flux in our analysis
period (e.g.,\citep{Ramazani19}).
1ES\,1218+304 is a TeV-detected HBL object at a redshift of $z=0.182$ that was determined by the spectroscopy of its host galaxy \citep{bade98}.
Its VHE emission above 120 GeV was first detected by the MAGIC (Major Atmospheric Gamma ray Imaging Cherenkov) telescope in
2005 \citep{albert06}. In May 2006, 1ES\,1218+304 was the target of HESS (High Energy Stereoscopic System) observation campaign and
these observations did not yield any statistically significant signal from the source \citep{aharonian08}.
The first evidence for the variability in VHE emission was detected by the VERITAS Observatory during the high activity of the source
in 2009 \citep{acciari10}.
The Fermi-LAT (Large Area Telescope) reported this source is one of the blazars with hardest spectrum above 0.1 GeV \citep{acero15,ajello17,nolan12}.

The organization of this paper is as follows.
Section 2 presents the multi-bands monitors and data reduction. The extracted light curves and analysis are described in Sections 3 and 4, respectively.
A brief discussion is presented in Section 5.



\section{OBSERVATION AND DATA REDUCTION}

On January 3, 4 and 5, 2019, the VERITAS Observatory found that 1ES\,1218+304 has a flare in VHE Gamma-ray.
As follows, the source was observed by \it Swift \rm for the next 4 days, which shows that there is no sign of decline of flux.
So, the source was continually observed in the \it Swift \rm mode until January 14.

At the same time, we monitored the object in multi-bands by using the GWAC-F60A telescope
at Xinglong Observatory of National Astronomical Observatories, Chinese Academy of Sciences (NAOC) during January 7 to 14, 2019.
The telescope with a diameter of 60cm is founded by Guangxi University, and operated jointly by NAOC and Guangxi University.
The telescope is equipped with an $1024\times1024$ iXon Utra 888 of EMCCD mounted at the Cassegrian
focus with a focal ratio of $f/8$.
The field-of-view of the telescope is 19 arc minutes.

The standard Johnson-Bessell $B$- and $R$-bands are used in our monitor.
The exposure time is 120 s in both bands. The typical seeing is $2{''}-3{''}$ during our observations.
In total, we obtained 367 and 371 images in the $R$- and $B$-bands in the eight nights, respectively.

A dedicated pipeline is developed by us to reduce the raw data by following the standard routine in the IRAF\footnote{IRAF is distributed by the National Optical Astronomical Observatories,
which are operated by the Association of Universities for Research in
Astronomy, Inc., under cooperative agreement with the National Science
Foundation.} package, including bias subsection, flat-field correction and subsequent aperture photometry.
In the pipeline, the readout noise and gain of each image are determined by using the \it findgain  \rm task.
In order to taking into account the effect due to variable seeing,
the aperture radius and the corresponding radius of the sky annuli are determined from a measurement of point-spread-functions of the bright field stars for each frame.
Because there is a close star next to the target, in order to avoid the contamination caused by the star,
we adopt a relatively small photometric aperture of $1.2\times $FWHM,
and a large inner radius of sky annuli of $5\times $FWHM for both target and reference star.
After standard aperture photometry, absolute photometric calibration is carried out based upon two nearby comparison stars whose magnitudes in the
Johnson-Cousins system are transformed from the SDSS Data Release 14 catalog through the Lupton (2005) transformation.
The 1$\sigma$ uncertainty in brightness is typically below 0.02 mag in both bands.

\section{Results and Analysis}
\begin{figure}
\centering
\includegraphics[width=13cm]{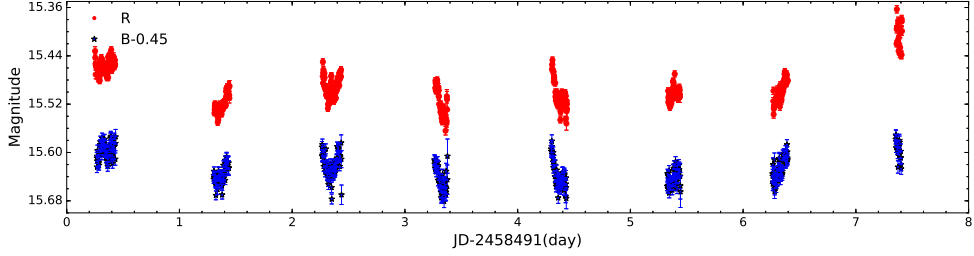}
 \caption{Light curves of the object in $B$, $R$band}
 \label{1}
\end{figure}

\begin{figure}
\centering
\includegraphics[width=6cm]{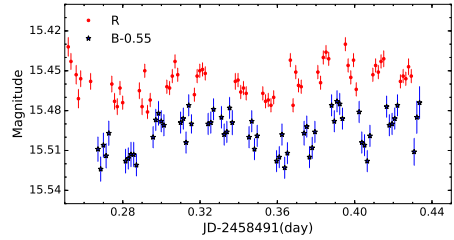}
 \caption{A possible QPO in the light curves in both $B$ and $R$ bands on JD2458491. }
 \label{2}
\end{figure}

The resulted light curves in both $B-$ and $R-$bands are displayed in the upper panel of Figure~\ref{1}.
At fitst glance, one can clearly see an intra-day varibility (IDV) for both light curves, although the object is lack of
evident long-term (i.e., days) variability during our monitors. In JD2458491, there is a possible
quasi-period oscillation in both bands, which has been reported occasionally in previous in other objects \citep{poon09,wu05},
although a quantified conclusion is failed because of the short sampling duration.
The light curve of the first is shown in Figure~\ref{2} for visibility.

\subsection{Variability Test}

Two statistical methods, a $\chi^2$-test and a ANOVA test \citep{diego10,Diego98}, are adopted to confirm and study the IDVs in the light curves.
In the $\chi^2$-test, the value of $\chi^2$ is determined for each day as follows:
\begin{equation}
    \chi^2 = \sum_{i=1}^N \frac{(V_i - \overline{V})^2}{ \sigma^2_i},
\end{equation}
where $V_i$ is the magnitude of the $i$th image, $\sigma_i$ the corresponding error, and
$\overline{V}$ the mean magnitude of all $N$ images.
A scaling factor of 1.5 is used to estimate $\sigma_i$ from the photometric error reported by the IRAF/phot task,
because the error is found to be underestimated by a factor of 1.3 to 1.75 by the phot task \citep{Gupta08, agarwal15}.
An IDV is positively detected if the calculated $\chi^2$ is larger than the corresponding critical value that is obtained from the standard $\chi^2$ distribution.


In the ANVOA test, we first divide the daily data into $k$ groups each with $n$ data points.
The inter-group ($SS_{\rm R}$) and intra-group ($SS_{\rm G}$) deviations can then be calculated as
\begin{equation}
    SS_{\rm G} = \sum_{j=1}^{k} (y_i -\overline{y})^2,
    SS_{\rm R} = \sum_{j=1}^{k} \sum_{i=1}^{n_j} (y_{ij} -\overline{y_j})^2,
\end{equation}
where $y_j$ and $\overline{y}$ are the mean values of the $j$th group and the whole data, respectively. A $F$-value is therefore
determined through
\begin{equation}
   F= \frac{SS_{\rm G}/(k-1)}{SS_{\rm R}/(N-k)}.
\end{equation}
Again, a positive detection of an IDV is returned if the calculated $F$-value is larger than the corresponding critical one.

The results of both tests are presented in Table~\ref{table:test result}. For both tests, a positive detection of IDV is denoted
by Y, and a negative detection by N (Columns (6) and (9)).
In conclusion, the consistence of the two tests enables us to confirm a positive detection of IDV in both bands in five nights, and
a negative detection in $B$-band in JD2458496. We refer the readers to \citep{singh20} for a discussion of the reasons
why the two tests may be different, and of the advantage and drawback of each test.

%

\begin{table*}
\centering
 \renewcommand{\arraystretch}{1.2}
  \caption{Results of ANOVA test and $\chi^2$-test of light curves in
  each band for each day\label{table:test result}}
  \begin{tabular}{@{\extracolsep{8pt}}cccrcccrcc@{}}
  \hline
  JD&Band&$N$&&\multicolumn{2}{c}{$\chi^2$-test}&& \multicolumn{2}{c}{ANOVA test}\\
  \cline{4-6} \cline{7-9}
  & & & $F$&$CV$&$Var?$& $F$&$CV$&$Var?$&\\
  (1)&(2)&(3)&(4)&(5)&(6)&(7)&(8)&(9)\\
 \hline
 2458491&$B$&60&75.8&87.1&N&4.6&3.5&Y\\
 2458491&$R$&61&134.7&88.3&Y&3.9&3.5&Y\\
 2458492&$B$&49&190.2&73.6&Y&10.8&4.3&Y\\
 2458492&$R$&50&201.7&74.9&Y&26.1&4.3&Y\\
 2458493&$B$&62&304.9&89.6&Y&5.1&3.5&Y\\
 2458493&$R$&60&309.2&87.2&Y&13.6&3.5&Y\\
 2458494&$B$&40&208.3&62.4&Y&9.0&5.1&Y\\
 2458494&$R$&40&299.3&62.4&Y&21.0&5.1&Y\\
 2458495&$B$&50&262.4&74.9&Y&23.1&4.3&Y\\
 2458495&$R$&50&441.8&74.9&Y&20.1&4.3&Y\\
 2458496&$B$&41&100.2&63.7&Y&1.9&5.1&N\\
 2458496&$R$&38&92.5&59.9&Y&8.5&5.1&Y\\
 2458497&$B$&50&97.0&74.9&Y&13.0&4.3&Y\\
 2458497&$R$&48&210.5&72.4&Y&22.4&4.3&Y\\
 2458498&$B$&19&25.8&34.8&Y&6.3&14.0&N\\
 2458498&$R$&20&111.9&36.2&Y&0.6&14.0&N\\
 \hline
\end{tabular}
\end{table*}

\section{ Relation of Color and Magnitude}

\begin{figure}
\centering
\includegraphics[width=15cm]{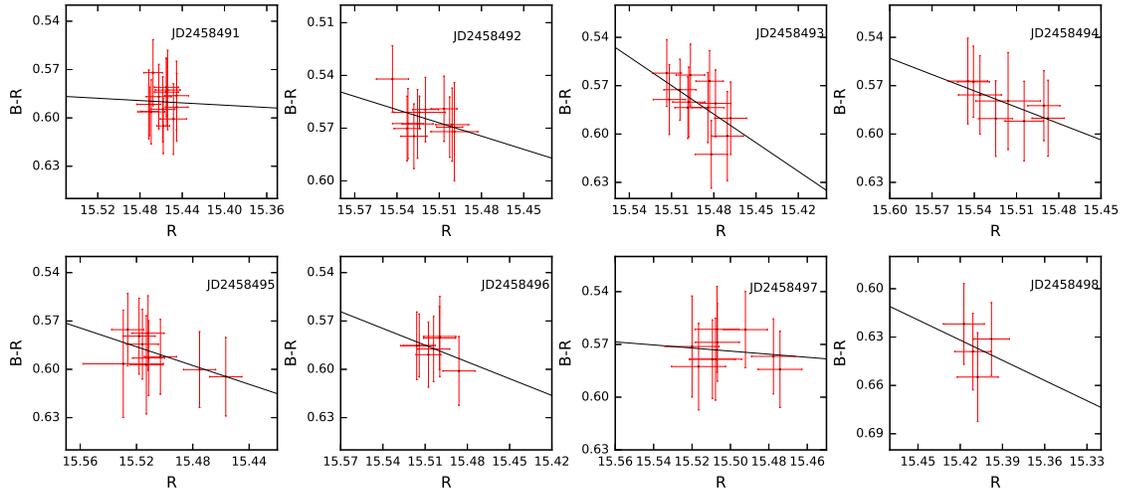}
 \caption{Color index of 1ES1218+304 in $B$ and $R$ band vs $R$ band every day,the black solid line represents the linear fitting.}
 \label{3}
\end{figure}

Figure~\ref{3} displays the relationship between the $B-R$ color and $R-$band brightness for each day.
Taking into account of the potential time delay between $B$ and $R-$band light curves, we calculate the $B-R$ colors after re-sampling the $B$ and $R-$band light curves with the same time sampling by binning the light curves.
To quantify the relationships, we perform a linear fitting to the observed data and calculate the corresponding Pearson correlation coecffients $r$.
The resulted slopes, $r$ and result of color behavior are listed in the Columns (2), (3) and (4) in Table~\ref{table:color}, respectively.

BWB chromatism can be clearly identified from the second night to the sixth night.
However, the variations are achromatic in the other nights.
On the one hand, BWB chromatism on short timescale has been frequently revealed in Blazars
in an active or flaring state \citep{wu06}, which is consistent with the shock-in-jet model \citep{dai15}.
In the model, when the jet strikes a region of high electron population as the shock propagates down, radiations at different frequencies
are produced at different distances behind the shocks.
High-frequency photons from synchrotron mechanisms typically emerge sooner and closer to the shock front than the low-frequency photons do,
thus causing color variations \citep{agarwal15}.
On the other hand, we argue that the observed achromatic variation is hard to be understood because of the lack of QPO, based on the models proposed in literature.
Please see Section 6 for more details.

\begin{table*}
\centering

\renewcommand{\arraystretch}{1.2}
  \caption{fit coefficient\label{table:color}}
  \begin{tabular}{@{\extracolsep{10pt}}cccc@{}}
  \hline
  JD&$A$&$r$&result\\
 \hline
 (1)&(2)&(3)&(4)\\
 2458491 & $-$0.036  &0.35 &NO\\
 2458492 & $-$0.251   &0.73 &BWB\\
 2458493 & $-$0.591  &0.44 &BWB\\
 2458494 & $-$0.338  &0.71 &BWB\\
 2458495 & $-$0.291  &0.73 &BWB\\
 2458496 & $-$0.346  &0.85 &BWB\\
 2458497 & $-$0.088  &0.49 &NO\\
 2458498 & $-$0.418  &0.39 &NO\\
 \hline
\end{tabular}
\end{table*}

\section{Cross-correlation and Time lags}

Diverse methods have been developed to analysis time delays between light curves in different wavelengths. The methods include
the discrete correlation function (DCF) method \citep{Edelson88},
the interpolated cross-correlation function (ICCF) method \citep{gaskell87}, and
the $z$-transformed discrete correlation function (ZDCF) method \citep{alexander97}.
We here use the ZDCF to analysis the time lag between $B$- and $R$-bands.
It has been shown in practice that the calculation of the ZDCF is more robust than that of the ICCF and the DCF when applied to sparsely and
unequally sampled light curves \citep{edelson96,giveon99,roy00}.
The ZDCF corrects several biases of the discrete correlation function method proposed by Edelson \& Krolik (1988) by using equal population binning
and Fisher's $z$-transform. \citep{Alexander13} uses ZDCF to
1) uncover a correlation between AGN's magnitude and variability time scale in a small
simulated sample of very sparse and irregularly sampled light curves through auto-correlation function; and 2)
estimate the time-lag between two light curves from a maximum likelihood function for the ZDCF peak location.

\begin{figure}
\centering
\includegraphics[width=15cm]{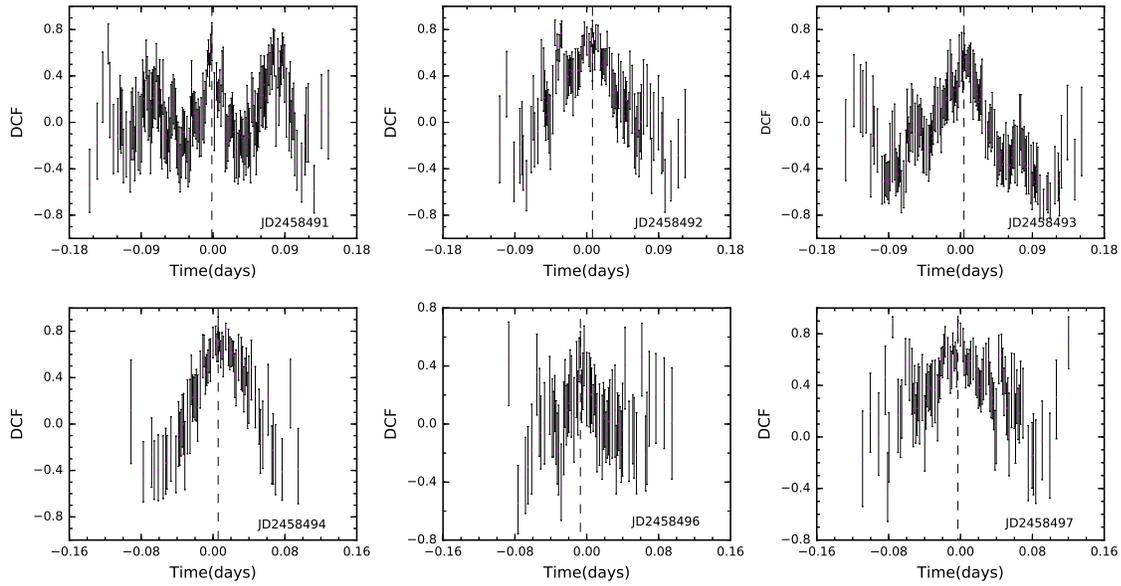}
 \caption{The ZDCF of 1ES1218+304 between $B$ and $R$ band. In each panel, the vertical black dashed line marks peak of ZDCF after a  ML estimation.}
 \label{4}
\end{figure}

Fig~\ref{4} shows the intre-band ZDCF for the optical light curves.
We use maximum likelihood (ML) estimation \citep{Alexander13} to determine the peak of ZDCF, then get the time lags.
The results of the ML estimation are tabulated in Table~\ref{zdcf},
where Column (1) is the Julian date, Column (3) is the peak of ZDCF and Columns (4) is the time lags and the corresponding ML interval.
The results show a positive lag in three nights, and a negative lag in three nights.
In JD2458495, there is no lag, because the value is less than one minute.

\begin{table}
\centering

\renewcommand{\arraystretch}{1.8}
\caption{Result of peak of ZDCF and Time lags\label{zdcf}}
\label{lags}
\begin{tabular}{ccccc}

\hline
JD&Passands &\multicolumn{2}{c}{ZDCF-ML}\\
&&ZDCF$_\mathrm{peak}$&Lag (day)\\
\hline
(1)&(2)&(3)&(4)\\
 2458491&$B-R$&0.71&$-0.001\substack{+0.001\\-0.004}$\\
 2458492&$B-R$&0.52&$0.007\substack{+0.015\\-0.033}$\\
 2458493&$B-R$&0.63&$0.004\substack{+0.014\\-0.002}$\\
 2458494&$B-R$&0.85&$0.006\substack{+0.016\\+0.001}$\\
 2458495&$B-R$&0.88&$<0.001\substack{+0.003\\-0.002}$\\
 2458496&$B-R$&0.52&$-0.007\substack{+0.07\\-0.008}$\\
 2458497&$B-R$&0.83&$-0.003\substack{+0.002\\-0.017}$\\
 \hline
\end{tabular}
\end{table}

\section{Conclusion and Discussion}

We monitored the BL Lac object 1ES\,1218+304 from January 7 to 14, 2019 in $B$- and $R$-bands, after the target was triggered by a VHE gamma-ray flare
detected by the VERITAS Observatory on January 3, 4 and 5, 2019.
The \it Swift \rm follow-ups indicates the object has the highest X-ray flux in history without clear decreasing.
Both ANVOA and $\chi^2$-test enable us to clearly reveal an IDV in optical in seven out of the eight nights.


A BWB chromatic relationship has been identified in five out of the eight nights, and achromatic variation in other nights.
The BWB behavior is usually observed in BL Lac objects, which can usually be explained with shock-in-jet models \citep{Zhang15}.
More therotical studies are need for understanding the lack of periodic phenomenon in the observed achromatic variation,
although several mechanisms have been in fact proposed to explain achromatic variation,
such as the lighthouse effect, geometrical effects and gravitational microlensing effects.
Achromatic variation is probably caused by the lighthouse effect in shock-in-jet model \citep{camenzind92}:
when enhanced particles move relativistically towards an observer on helical trajectories in the jet, flares will be produced
by the sweeping beam whose direction varies with time \citep{Man16}.
But the resulted light curve is periodic and achromatic, such as for S5076+714 \citep{Man16}.
The geometrical effects are that the helical or precessing jet leads to varying Doppler boosting towards the observers and hence the flux variations,
which is usually produce achromatic and periodic variations again \citep{Wu07}.
Gravitational microlensing effects is from external origin,
which usually generates the light curve of achromatic and symmetric.

A cross-correlation and time delay analysis allows us to find a positive delay between the $B$- and $R$-bands in three nights, and a negative one
in three nights. The minimum time delay is about 1.5 minutes, and the maximum one more than 10 minutes.

The mass of SMBH ($M_{\rm BH}$) in a Blazar can be  estimated by \citep{gupta12}
\begin{equation}
   M_{\rm BH}=\frac{c^3\Delta t_{\rm obs}}{10G(1+z)}.
\end{equation}
where $\Delta t_{\rm obs}$ is the observed shortest variability timescale, and $z$ is the redshift.
By equalling the minimum zero-crossing time of the ZDCF (in autocorrelation mode) to the shortest variability timescale approximately \citep{Meng18}, the
$M_{\rm BH}$ is calculated to be 2.8$\times10^8  \rm M_{\odot}$ for 1ES\,1218+304.
If the variations arise in the jets and are not explicitly related to the inner region of the accretion disc, the BH mass estimation is invalid \citep{Meng17}.



%



\normalem
\begin{acknowledgements}
The authors thank the anonymous referee for a careful review and helpful suggestions that improved the manuscript.
The authors are thankful for support from the National Key R \&D Program of China(grant No. 2020YFE0202100).
The study is supported by the National Natural Science Foundation of China under grant 11773036, and by the Strategic Pioneer Program on Space Science,
Chinese Academy of Sciences,grants No. XDA15052600 and XDA15016500. J.W. is supported by the Natural Science Foundation of Guangxi (2018GXUSFGA281007 \& 2020GXNSFDA238018), and by the Bagui Young Scholars Program.
Special thanks go to the night asistants of the GWAC system for their help and support in observations.

\end{acknowledgements}

\bibliographystyle{raa}

\end{document}